# Tile Optimization for Area in FPGA based Hardware Acceleration of Peptide Identification

S. M. Vidanagamachchi*, S. D. Dewasurendra*, R. G. Ragel* and M. Niranjan+

*Department of Computer Engineering,
Faculty of Engineering, University of Peradeniya
Peradeniya, Sri Lanka

+School of Electronics and Computer Science
University of Southampton
United Kingdom

*Abstract*— **Advances in life sciences over the last few decades have lead to the generation of a huge amount of biological data. Computing research has become a vital part in driving biological discovery where analysis and categorization of biological data are involved. String matching algorithms can be applied for protein/gene sequence matching and with the phenomenal increase in the size of string databases to be analyzed, software implementations of these algorithms seems to have hit a hard limit and hardware acceleration is increasingly being sought. Several hardware platforms such as Field Programmable Gate Arrays (FPGA), Graphics Processing Units (GPU) and Chip Multi Processors (CMP) are being explored as hardware platforms. In this paper, we take an FPGA hardware exploration and expedite the design time by a design automation technique. Further, our design automation is also optimized for better hardware utilization through optimizing the number of peptides that can be represented in an FPGA tile. The results indicate significant improvements in design time and hardware utilization which are reported in this paper.**

## I. INTRODUCTION

Deoxyribonucleic acid (DNA) is a nucleic acid that contains the genetic information of all known living organisms and is organized as chromosomes and genes reside on it. Genes are working subunits of DNA that carry the genetic information. These genes/DNA are used for protein synthesis in ribosomes and proteins are biochemical compounds consisting of one or more polypeptides (a single linear chain of amino acids bonded together by peptide bonds). The shortest peptides are dipeptides consisting of 2 amino acids joined by a single peptide bond.

Protein identification can be done through different ways and it is the initial step of submitting proteomics data to biological databases. Mass Spectrometry (MS) and Peptide Mass Fingerprinting (PMF) are two methods of protein identification. Mass spectrometry is an analytical method used to measure the molecular mass of a sample and can be used for understanding the chemical structures of molecules (peptides and other chemical compounds). Mass spectrometry can be of two main types; MALDI (Matrix-assisted laser desorption/ionization) peptide spectrum and Tandem peptide spectrum. Peptide mass fingerprinting is also an accurate, analytical technique for protein identification where unknown proteins are cleaved into small peptides and then their absolute masses are accurately measured by a mass spectrometer Next paragraph describes how string matching is used in identifying proteins/genes and this paper presents a way of identifying already known peptides in any given protein (known/unknown) sequence.

String matching is a classical problem and it is fundamental to many applications that need processing of text data or some sequence data. String matching has been widely studied in the past three decades. Several string matching algorithms are used to find and locate one or several string patterns that are found within a larger string/text. String matching algorithms are used not only in applications such as text editors, word processing and bibliographic search, but also in comparing biological sequences in bioinformatics with the remarkable increase in the number of DNA and protein sequences been identified. In computational biology exact string matching is commonly required. For example, in proteogenomic mapping proteomics data is used for genome annotation (identifying locations and coding regions of the genes in genome and determine their functionality). Here the result of mass spectrometry (identified peptides) is matched with the target genome (already identified genomes in databases) which is translated in all six open reading frames (DNA sequences which does not contain a stop codon) [2]. Exact string matching can also be used for sequence analysis. Pair wise sequence matching, multiple sequence matching, global alignment and local alignment are some types of string matching which are used in bioinformatics and computational biology.

String matching algorithms can be basically classified into three groups; single pattern matching algorithms (search a single pattern within the text), algorithms that uses finite number of patterns (search finite number of patterns within the text) and algorithms that uses infinite number of patterns (search infinite number of patterns within the text such as regular expressions). Single pattern matching algorithms include Rabin-Karp [24], finite state automation based search [25], Knuth-Morris-Pratt [26] and Boyer- Moore [27]. Rabin-Karp algorithm can be used for both single pattern matching and for matching a finite number of patterns. For Rabin-Karp, in a string of length n, if p patterns of combined length m are to be matched, the best and worst case running time are O (n+m) and O (nm) respectively. Finite state automata based search is expensive to construct (power-set construction), but very easy to use. Knuth-Morris-Pratt (KMP) algorithm turns the search string into a finite state machine, and then runs the machine with the string to be searched as the input string. Running time of this algorithm is O (n+m) [1]. Algorithms used for finite number of patterns (multi patterns) include Aho-Corasick [28], Commentz-Water [29], Rabin-Karp and Wu-Manber. Aho-Corasick is a dictionary matching algorithm and matches all patterns at once. Aho Corasick algorithm provides a scalable solution to the string matching problem and it has computational complexity of O (m+k) where k is the total number of occurrences in the pattern strings in the text. (It has the worst-time complexity of O (n+m) in space O (m)).

FPGA (Field Programmable Gate Arrays) technology is an emerging technology for providing better hardware solutions for sequence comparison, protein/molecular structure comparison and large scale clustering than other Programmable Logic Devices (PLDs). While maintaining its performance in par with other hardware solutions (such as ASICs), it is programmable. Tile is a small area of FPGA consists of logic cells.

The objective of this paper is to present a process of automating the implementation of an exact string matching algorithm (Aho Corasick) on an FPGA and tile optimization for area in FPGA. Tile architecture consists of several logic elements and here we map peptides into tiles using Finite State Machines. In tile optimization we tried to add the maximum possible peptides to a tile using our algorithm.

We start by giving a brief introduction to hardware and software implementations of string matching algorithms in section II (Related Work). Then section III describes the problem definition. Section IV includes the description of algorithm automation and tile optimization algorithm. The results of the tile optimization algorithm, data collection and preprocessing details are discussed in section V. Section VI includes the discussion.

## II. RELATED WORK

There exist several software and hardware implementations of the string matching algorithms and they are described next.

Software implementations of Knuth Morris Pratt algorithm includes C and $C^{++}$ implementations with running time of O (m+n) [5][6]. Aho Corasick algorithm has $C^{++}$ [7] and $C^{\#}$ [8] software implementations with running time O (n).

It is found that 31% of the Snort [32] processing (for intrusion detection) is due to string matching. Therefore for the efficiency of Snort, efficient string matching algorithms should be developed. Coit has proposed a string matching algorithm (software based approach) based on both Boyer Moore and Aho Corasick algorithms that can improve the performance of Snort by 1.02 to 3.32 times when comparing to the standard Boyer Moore implementation. Wu Mander multi pattern matching algorithm and E2xB are other algorithms implemented in Snort [9].

Hardware implementations of Knuth Morris Pratt Algorithm include Cyclone II FPGA implementation of NIOS II processor. The Nios II was responsible for ferrying data from SDRAM to the different search units and reading their results sequentially. Further it was also responsible for precomputing the table. In search units they have used data such as the database, string and table; therefore some sort of storage is needed to be used in the search units. Since this need parallel access to the data in every cycle SRAM or SDRAM couldn't be used. Instead they have used M4K blocks and each search unit uses 3 modules of this data memory. Initially they tested with 8 blocks on the FPGA (When more blocks are used compile time increases). Some constraints include length of the search string, which is 255 bytes, each database entry is 255 bytes and the full database size is 7MB. The database entries are also limited to 255 bytes. They have suggested some optimizations that could be made including increasing the number of blocks, RAM sizing and running the NIOS II code out of internal memory. By using bit splitting method to search larger strings, the memory space occupied be could be reduced. This memory reduction method is not implemented for this algorithm [10].

Yamani et al. [11] have done a FPGA design of Boyer Moore Algorithm for spyware detection in 2010. However it is not implemented yet.

The space and performance of Aho Corasick algorithm can be optimized by implementing it in hardware such as FPGA. Simplified version of Aho Corasick algorithm was used by Brundo to identify anchor points in CHAOS algorithm [23] for fast alignment of huge genomic sequences. Farre used Aho Corasick algorithm for predicting transcription binding sites in PROMO v.3 tool [24]. Then Hyyro found that Aho Corasick out performs other algorithms for locating unique oligonucleotides in the yeast genome [2]. For applications of Aho Corasick other than bioinformatics, Snort programme can be taken. Snort is a popular computer security programme that looks for a set of "signature" patterns corresponding to known intrusion attacks in network packets. Bit split Finite State Machine (FSM) implementation of Aho Corasick algorithm is more efficient in terms of hardware area than FSM implementation without splitting bits. Additional savings in the storage can be obtained by bit splitting implementation. Here it is achieved by splitting the FSM into smaller FSMs. In this bit splitting method, a single state machine is split into multiple machines each handles some fractions of the input string. The number of out edges per state is significantly reduced by splitting an Aho Corasick state machine into several state machines. Each state machine is responsible for a subset of the input bits; this will increase the number of states proportionately to be active in the system. Bit-split method removes most of the wasted edges in the search tree; therefore the required storage is smaller than the starting state machine [4]. Advantages of bit split method include; its maintain ability of the Aho Corasick machine to match strings in parallel and for each state it reduces the memory required for state transition storage [4]. This was implemented on the Xilinx Virttex-4 FX-100 FPGA, which consists of 376, 18-kbit BRAM blocks, of which 350 are used for implementing Aho-Corasick tiles and the remaining 26 are reserved for meeting the storage requirements of other modules.

A technique for improving table (includes state transitions) compression is described by Y. Liu in 2009 [12]. This method reduces huge memory usage of extended Aho corasick. Extended Aho Corasick automation refers to the full Aho Corasick automation that has eliminated failure transitions. Therefore it can be applied to large scale pattern sets. Here they have presented a simple and efficient table compression method to reduce the automation's space [12].

In 2008 Fei Xia and his group [13] have proposed a systolic array approach to detect string matches without using looking up tables. To accelerate first two stages of NCBI BLAST family algorithm they have implemented pipelining systolic array as a multi seed detection and parallel extension pipeline engine. This implementation consumes less memory resources and it has better performance results in both of processing element number and clock frequency accelerations. According to the results speedup could reach about 17, 48, 14, 71 and 10 compared to the NCBI BLASTp, TBLASTn, BLASTx, TBLASTx and BLASTn programs 3072 residue queries on Intel P4 CPU respectively [13].

In 2009 first FPGA implementation of the Position Specific Iterated BLAST algorithm was released. This implementation is parameterized in terms of sequence length, scoring matrix, gap penalties and threshold values. This implementation consists of several blocks and each of them performs one step of the algorithm in parallel [14].

Due to computational complexity, when running on general purpose computers, performing Smith Waterman algorithm is impractical for large databases. In 2004 Stefan et al. [15] have found a memory efficient significantly accelerated FPGA implementation for smith waterman algorithm. They have proposed a different parellization scheme than commonly used for systolic arrays. This leads to full utilization of processing units regardless of sequence length. FPGA implementation of Smith-Waterman algorithm can accelerate the alignment by two orders of magnitude on a Pentium desktop comparing to standard OSEARCH program (an alternative version of the SSEARCH programme) [16].

In 2007 Isaac et al. [17] and Peiheng et al. [18] found two FPGA implementations of Smith- Waterman algorithm. In the first implementation they have discovered a method for accelerating Smith Waterman (SW) algorithm using FPGA that implemented a module to compute the score of a single cell of the SW matrix. Then through the FPGA circuit entire SW matrix was computed (using a grid of the above module). This method gradually accelerate the computing time by up to 160 folds compared to a pure implementation running on the same FPGA with an Altera Nios II softprocessor [17]. In the second implementation they presented implementation of Smith Waterman algorithm for both DNA and protein sequences. It includes a multistage processing element (PE) design which allows more parallelism and reduces the FPGA resource usage, a pipelined control mechanism, a compressed substitution matrix storage facility and a key to minimize the overall PE pipeline cycle time. This implementation results in acceleration of 185 and 250 compared with the 2.2 GHz AMD Opteron host processor [18].

In 2007 Fei Xia and Yong Dou [19] have found a storage optimization method for hardware accelerating Needleman Wunsch [31] algorithm. This optimized implementation stores a part of the scoring matrix and it reduces the storage usage of FPGA RAM blocks and implements more processing elements in FPGA. The results show that the peak performance can reach 77.7 GCUPS (Giga cell updates per second) and 46.82 GCUPS respectively [19].

ClustalW [30] multiple sequence alignment tool consumes too much time to perform on state-of-the-art workstations. Accelerated method for performing ClustalW using reconfigurable hardware was introduced in 2005 by Tim et al. [20].

Istvan et al. [21] proposed a powerful solution to process real time mass spectrometric data generated by MALDI-TOF instruments. This implementation with de-noising, baseline correction, peak identification and deisotoping running on a Xilinx Virtex 2 FPGA at 180 MHz produces a mass fingerprint over 100 times faster (almost 170 fold speed gain relative to a conventional software running on a dual processor server) than an equivalent algorithm written in C running on a Dual 3 GHz Xeon workstation [21].

Istvan et al. [22] presented a parallel database search engine for Peptide Mass Fingerprinting; it delivers 1800 fold speed up when running on a Xilinx Virtex 2 FPGA at 100 MHz compared with an equivalent C software routine. This implementation provides a complete real time PMF protein identification solution. This was implemented and tested consisting of a FPGA motherboard equipped with a Xilinx Virtex- II XC2V8000 FPGA (consists of 8 million gates) and 4 MB RAM, communicate with the host PC server through a PCI (Peripheral Component Interconnect) interface. The implementation consists of three FPGA modules, Mass Spectrum processor and a PC server. Only one FPGA was used to implement the database search engine. This search processor performs two basic operations; simulation for protein digestion with peptide mass calculation and matching score calculation. Further this search engine occupies 99% of the FPGA's logic resources, 99% of the FPGA's internal RAM resources and 53% of the FPGA's I/O resources.

Since Aho Corasick algorithm has the best performance among multiple string matching algorithms we automated it in FPGA for peptide identification.

No details available on the design automation of Aho-Corasick Algorithm on FPGAs in the literature. Therefore, this is the first time such automation and related optimizations are reported.

III. PROBLEM DEFINITION

During last few decades, advances in life sciences have lead to the generation of a huge amount of biological data. Therefore there is a pressing need for efficient computational methods to cope with them. However a significant bottleneck exists in the analysis of such data. According to the predictions of Moore's Law the number of transistors that can be placed on an integrated circuit has doubled approximately every two years, but computational demand for analyzing huge amount of biological data is growing faster than the increase in processing power of computers. Many attempts have been made by several research groups to develop efficient algorithms as well as dedicated hardware/software solutions to deal with this explosion.

Manual implementation of Aho-Corasick algorithm on an FPGA is challenging due to its time consuming nature. As there is no details of automating Aho-Corasick Algorithm implementation on FPGAs in the literature [1][2], we give some information of the automation we performed.

Bit split Aho Corasick algorithm can be used to accelerate peptide identification using FPGA. In Yoginder et al. [2] implementation of Aho Corasick in FPGA uses alphabetical order of peptides to build the Finite State Machines (FSMs). In their implementation one Finite State Machine consists of a maximum 20 peptides. For one tile in the architecture there are five bit split versions of FSMs to identify peptides. We re-implemented [2] and according to the results it is not the optimal order of adding peptides to FSMs. Apparently this is because of the order of the bit split versions of peptides. They are not sorted. In Yoginder et al. implementation they have limited the maximum number of peptides per a tile to 20 for the maximum number of 256 states. Therefore we increase the number of peptides per tile by changing the order of adding peptides in order to use a lesser number of logic cells in FPGA.

In the first instance we used a trial and error method to select the next peptide to be included as the last peptide in a tile (elaborated in Section IV-A). Even though we managed to increase the number of peptides per tile by this means, the increment is marginal. Therefore we implemented another algorithm (Algorithm 1) considering the length of peptides of bit spitted peptides. Even it is better than their algorithm there can be exceptional random order (there can be several peptides of same length) which gives better order to minimize the total number of states in the state machine.

## IV. METHODOLOGY

Since Aho-Corasick algorithm is the best and the widest used multiple pattern matching algorithm which searches all occurrences of any of a finite number of keywords in a text string, Yoginder et al. [2] have used this algorithm for hardware acceleration of peptides pattern matching for the 1$^{st}$ chromosome of human genome. This algorithm consists of two phases; constructing a finite state machine from keywords (Figure 1) and then using the state machine for locating the keywords by processing the text string in a single pass. They have used bit split implementation of Aho-Corasick to reduce storage space. Here each amino acid is represented by 5 bits and then 5 Finite State Machines (FSM) are generated for the given set of peptides. An FSM can be constructed as a graph or a keyword tree, which consists of several states. FSM creation starts with the initial state '0' and then according to the given peptide (keyword) input it goes to next states (for new incoming characters) and finally it creates the machine deterministically. Final state of a given peptide represents the matching state. This algorithm can be used only with exact string matching applications and cannot be used with approximate string matching. We have automated the implementation of the hardware system using the C$^{++}$ programming language.

Initially we developed the search tree according to the Aho Corasick algorithm. Then we created VHDL code, graphs and tables by traversing this tree.

Our software can generate a full FSM as well as five bit-split FSMs automatically for a given set of peptides. This system outputs each bit-split FSM representing a bit of each amino acid (since there are 20 total numbers of amino acids each amino acid can be encoded into 5 bits) as VHDL models. The software makes use of tables for each and every FSM, which contain the states. Columns of these tables represent possible amino acids and rows represent all possible states (therefore data in the table represent next state according to the relevant input amino acid). This automation software also creates graphs (using graphviz software http://www.graphviz.org/) for each finite state machine indicating every state.

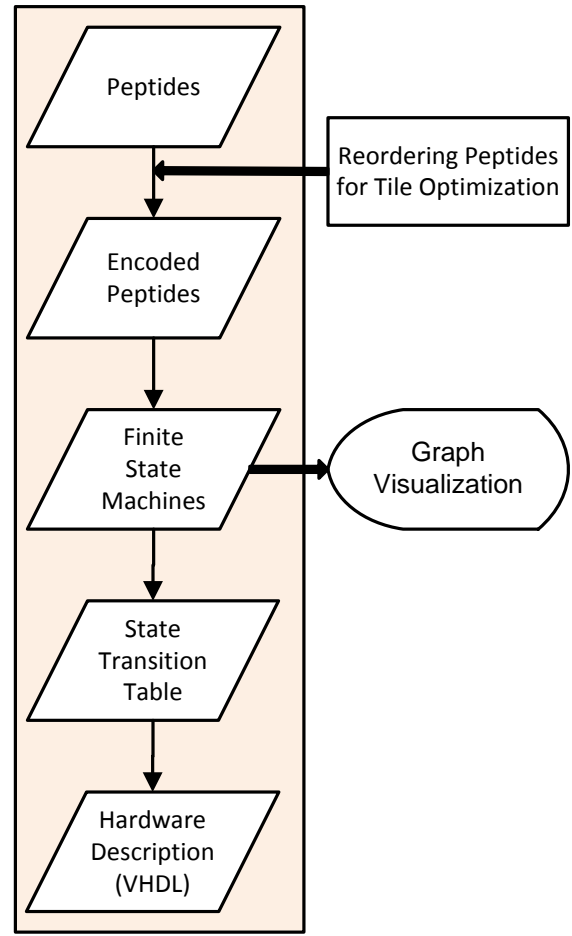

Figure 1. The process of optimization and Automation

The peptide reordering performed in order to optimize the tile utilization is of the peptides are discussed in the rest of this section.

## A. Increasing the number of peptides per tile by modifying Yoginder et al. implementation

When we add peptides in alphabetical order for a tile we can add until a maximum of 256 states are assigned. However, we may face a problem in adding the last peptide if its addition would require more than 256 states to a tile. We resolved this problem by adding next suitable peptide by searching through the rest of the peptide list (trial and error method) which is arranged according to the alphabetical order. After it is added to the tree (tile) we erase the peptide from the list.

## B. Changing the order of adding peptides (Algorithm): Steps

One of the major contributions of the optimization for better utilization of the tiles is described here. We started from the beginning with a list of peptides and followed the steps shown in Algorithm 1. The major steps of the algorithms are described below:

- Split all the amino acids of peptides into five bits (Each amino acid representing 5 bits: for eg:-A=00000, C = 00001, E = 00011 then corresponding 5 bit split sequences of peptide ACE are: 000,000,000,001 and 011).(Algorithm 1:Line 2)

- Started with a randomly selected peptide in the peptide pool. (Algorithm 1:Line 4)

- Add peptides: select the next best suitable peptide that results in the minimum number of states and add it to a list (Algorithm 1: Line 7, 18, 21, 30, 33, 44) (finally add them to the search tree/FSM).

- Repeat the 3$^{rd}$ step until all peptides get added to the FSM.

- Do above steps for each bit split peptide set and finally find the total number of minimum states for ach peptide set. (Algorithm 1: Line 21,34)

## V. DATA COLLECTION, PRE-PROCESSING AND RESULTS

We have used protein data from GenBank [34] database and PeptideMass [33] software which is available in Expasy Proteomics Server to generate peptide PeptideMass creates possible peptides for a given protein. Then these peptides are input in the software system (implemented in C++) to generate the VHDL (Very-high-speed integrated circuit Hardware Description Language) implementation. For example if we consider a tile which has the average length of 4 amino acids, maximum length of 8 amino acids and minimum length of 3 amino acids of peptides; manually writing a VHDL model (for hardware) to match 20 peptides takes about 2 hours for an experienced hardware designer. However, our automated system takes around 150- 200 milliseconds to do the same and 450 – 475 milliseconds to generate all FSMs, tables and graphs in an Intel (R) Core (TM)2 Duo CPU 2 GHz with 2GB RAM and 32-bit operating system. Therefore automation is 16000 times faster than the time taken by an experienced developer.

```
1   // Algorithm for selectincting the order of peptides
2   if(GetBitSplitString()!= 0){return 1}
3   if(c==1){
4     minPep = (*MyItr);MyItrMin = MyItr; min = sum_1;MyItr++;
5     if(MyItr == sListVar.end()){
6       if(min<=sum_1){
7         min = min;MyItrMin = MyItrMin; minPep = minPep;
8       }
9       else{
10        min = sum_1;minPep = (*MyItr);MyItrMin = MyItr;
11      }
12      sListPep.push_back(minPep);sumNodes = min+sumNodes;
13    }
14  }
15  else{
16    if (c == sListVar.size()){//reach the last peptide
17      if(min<=sum_1){
18        min = min;MyItrMin = MyItrMin;minPep = minPep;
19      }
20      else{
21        min = sum_1; MyItrMin = MyItr;minPep = (*MyItr);
23      }
24      sListPep.push_back(minPep); sumNodes
                                    = min+sumNodes;
25      sListVar.erase(MyItrMin); //delete the peptide from list
26      MyItr = sListVar.begin();c = 0;
27    }
28    else{
29      if(min<=sum_1){
30        min = min;MyItrMin = MyItrMin; minPep = minPep;
31      }
32      else{
33        min = sum_1; MyItrMin = MyItr; minPep = (*MyItr);
34      }
35      MyItr++;
36    }
37  }
38  c++;
39  }
```

Algorithm 1. Tile Optimization Algorithm

In our experiment after removing the limitation of 20 peptides (but maximum of 256 states) per tile and then finding the best possible next peptide to be added as the last peptide, we could add maximum number of 33 peptides per tile and the minimum was 8 peptides per tile (this is because of peptides with long lengths). Here we need only 96 tiles to map the entire set of 2800 peptides. Here total number of nodes in the tree which represents the bit split FSMs is 120923.

If we consider the length of peptides it is better than considering alphabetical order when selecting peptides and also sometimes there could be other random combinations that may have least number of total states (However even when we gave the peptides as inputs to the algorithm according to their

length there tree can be some other random order of peptides giving a better order as an exception)

For tile optimization we have used 11 sets of 250 numbers of peptides in alphabetical order and 10 sets of 250 numbers of peptides in random orders. Furthermore we performed the same for a 2800 numbers of peptides set once.

According to this algorithm we could gain around 16% increment of peptides per tile when considering total number of states (when adding 200 peptides from 250) than the alphabetical order. This is only for our data sets which are given in alphabetical order and randomly selected orders; if the length of the data set is small it should increase.

TABLE I.  TILE OPTIMIZATION RESULTS I

| No. of tiles | | |
|---|---|---|
| **Alphabetical** | **Ordered** | **Tile increment** |
| 96 | 68 | 29.1% |

As mentioned in an earlier paragraph we need 96 tiles to map entire set of peptides with alphabetical order. After we ordered the peptides set according to our algorithm it requires only 68 tiles with 33 maximum numbers of peptides per tile. Therefore we could gain 29.1% tile increment (Table I) using this method for our data set and 70.6%, 43.3%, 25.1%, 9.28% peptide increment for 65175, 98200, 134570, 169820 total number of states respectively when adding all peptides to one tile (Table II).

TABLE II.  TILE OPTIMIZATION RESULTS II

| Total # of States | Number of Peptides | | |
|---|---|---|---|
| | **Alphabetical** | **Ordered** | **Pep increment** |
| 65175 | 1000 | 1706 | 70.6% |
| 98200 | 1500 | 2150 | 43.3% |
| 134570 | 2000 | 2502 | 25.1% |
| 169820 | 2500 | 2732 | 9.28% |

## VI. CONCLUSION

It is time consuming and tedious to manually write a VHDL model that matches a large number of peptides. Therefore this automation gives an efficient and convenient way of implementing hardware in VHDL by just specifying the set of peptides. It is better to consider the length of the peptides when adding peptides to the tiles than alphabetical order.